\documentclass[a4paper]{article}

\usepackage[USenglish]{babel} 
\usepackage[T1]{fontenc}
\usepackage[ansinew]{inputenc}

\usepackage{lmodern} 

\usepackage{graphicx} 

\usepackage{amsmath}
\usepackage{amsthm}
\usepackage{amsfonts}

\begin{document}

\pagestyle{empty} 


\title{Generating functions for anti-canonical transformations  in the Zinn-Justin and Batalin-Vilkoviski formalisms}


\author{A Andra\v si$^+$ and J C Taylor\footnote{Corresponding author } 
\footnote{\textit{E-mail addresses} aandrasi@irb.hr (A. Andrasi), jct11@cam.ac.uk (J. C. Taylor)} \\ \\ {\it  $^+$Vla\v ska 58, Zagreb, Croatia} \\ $^\dagger${\it DAMTP, Cambridge University, UK}}





\maketitle



\begin{abstract}

\noindent{Quantization of gauge fields by the BRST method requires sources in addition to fields, and a 
bilinear anti-bracket defined in terms of them. This bracket is a sort of generalization
of a Poisson bracket in classical mechanics. Canonical transformations are also generalized as
anti-canonical transformations. In this paper, we take the analogy with classical mechanics
one step further, by showing how anti-canonical transformations can be derived from generating functions.
We give an example relevant to the renormalization of QCD in the Hamiltonian formalism.}\\

\noindent{Pacs numbers: 11.15.Bt; 03.70.+k}\\

\noindent{Keywords  gauge fields, renormalization}

\end{abstract}

\vfill\newpage

\pagestyle{plain} 

\section{Introduction}
The BRST method \cite{BRS} for quantizing gauge fields involves a source for the increment of each field (including the DeWitt-Faddeev-Popov ghost) under an infinitesimal gauge transformation. A bilinear star operation
of any two functionals, $F*F'$, is defined in terms of these fields and sources, with the action, $\Gamma$
satisfying $\Gamma * \Gamma=0$.

Zinn-Justin \cite{ZJ} and Batalin and Vilkoviski \cite{BV} introduced  compact notations for the fields and sources, which we will call
$q_i, p_i$ (anticipating an analogy with classical mechanics). For each $i$, just one of $q_i$ and $p_i$ is a field and the other is  source, and just one is bosonic and the other is fermionic (that is Grassmann odd). In Zinn-Justin's formulation, all $q_i$ are bosonic and all $p_i$ fermionic.
In Batalin and Vilkoviski's formulation, all $q_i$ are fields (including the ghost field) and all $p_i$ are sources. The suffix $i$ includes discrete indices like Lorentz and colour indices, and also in general the spacetime coordinate $x$.  

Batalin and Vilkoviski \cite{BV} (for reviews see \cite{W} \cite{G}) have generalized the $*$ operation
by a bracket $(F,F')$, between two functionals $F$ and $F'$of $q,p$  which may be fermionic  as well as bosonic. Their definition is 
\def\d{\delta}
\begin{equation}
  (F,F')= \frac{\d_R F}{\d q_i}\frac{\d_L F'}{\d p_i}-\frac{\d_R F}{\d p_i}\frac{\d_L F'}{\d q_i},
\end{equation}
where a sum (and integral) over $i$ is understood, and the suffices $L$ and $R$ denote left- and right-differentiation.  In Zinn-Justins's formulation, it would not be necessary to use right-differentiation as well as the normal left-differentiation, but we will use the more general definition (1). The only thing to note about right-differentiation is that the right-differential of a bosonic quantity by a fermionic quantity is the negative of the left-differential.

We can label the $q_i$ and $p_i$ so that $q_i$ is bosonic for $i=1,....m$ and fermionic for $i=m+1,...n$
and the $p_i$ oppositely ($m=n$ in the Zinn-Justin formalism). If $F$ is bosonic and $F'$ is either bosonic or fermionic, equation (1) gives
\begin{equation}
(F,F')=\sum_1^m\left[\frac{\d F}{\d q_i}\frac{\d F'}{\d p_i}+\frac{\d F}{\d p_i}\frac{\d F'}{\d q_i}
\right]
-\sum_{m+1}^n\left[\frac{\d F}{\d q_i}\frac{\d F'}{\d p_i}+\frac{\d F}{\d p_i}\frac{\d F'}{\d q_i}
\right].
\end{equation}
(All derivatives are now left derivatives, unless stated otherwise.) Also $(F',F)=\pm(F,F')$ with the plus sign if both $F$ and $F'$ are bosonic and the minus sign otherwise. (For the Zinn-Justin case, we made a very similar definition to (2) in the appendix of \cite{AAJCT}, except that the sign was opposite 
for $F$ fermionic.)

The BRST method requires that the effective action, $\Gamma$, of  a gauge theory must satisfy
\begin{equation}
(\Gamma,\Gamma)=0.
\end{equation}
In view of (2), this requires that, \textit{if} $\Gamma$ is linear in the bosonic currents (such as the current coupled to the gauge transform of a ghost), $p_{m+1}...p_n$ must appear in $\Gamma$ with opposite signs in the Zinn-Justin and Batalin-Vilkoviski formalisms. In this case, changing the signs of $p_{m+1}....p_n$ in $\Gamma$ reduces the latter formalism to the former. Of course, the BRST transforms are then the same in both formalisms. In  Appendix 2, we give an example of what happens with
parts of $\Gamma$ which are nonlinear in the bosonic currents.

The bracket (1) has a superficial resemblance to a Poisson bracket. But it has modified properties. (1) is not  anti-symmetric under $F\leftrightarrow F'$ if both $F$ an $F'$ are bosonic.
It obeys modified Leibniz and Jacobi identities (see \cite{G}).

In analogy with the canonical transformations of classical mechanics, an anti-canonical transformation from the fields/sources $q_i,p_i$ to new ones $Q_i,P_i$ is defined to be one satisfying the conditions
\begin{equation}
(Q_i,P_j)=\delta_{ij}=-(P_j,Q_i),\,\,\,\, (Q_i,Q_j)=0,\,\,\,\, (P_i,P_j)=0,
\end{equation}
(where the $\delta_{ij}$ is interpreted to include a Dirac delta function of spacetime). We  assume further that if, for each $i$, if $q_i$ is bosonic/fermionic, then  so is $Q_i$, and similarly with $p_i$ and $P_i$. An anti-canonical transformation on the independent variables $q_i,p_i$ in (2) does not change its value. This is verified in Appendix 1.

In renormalization theory, the relation between bare and renormalized fields and sources is an example
(usually trivial) of an anti-canonical transformation.

To first order, an anti-canonical transformation can be generated as follows. Define
\begin{equation}
G(q,p)=q_ip_i+J(q,p),
\end{equation}
where $J$ is any functional proportional to some small coupling constant, say $g$. We also assume that $J$ is fermionic.
Then
\begin{equation}
Q_i(q,p)=\frac{\d G}{\d p_i},\,\,\,\,\,P_i(q,p)=-\frac{\d G}{\d q_i}
\end{equation}
satisfy (4) to first order in $J$. In renormalization theory, this corresponds to one-loop order,
and to go to higher order one proceeds iteratively. In (6) we have not distinguished between
left- and right-differentiation: because $G$ is fermionic, they would be the same. 

In classical mechanics
there are generators of all-order canonical transformations, using mixed functions $G(q,P)$ or $G(Q,p)$ etc. which implicitly define $Q(q,p), P(q,p)$. In the next section we prove that this is possible with anti-canonical transformations.
In section 3
 we demonstrate the existence of inverses of  matrices $M_R$ and $M_L$ defined in (9) below. In section 4, we show how, in the case of the Zinn-Justin formalism, to derive expansions for $Q,P$ in powers of $J(q,p)$ and its derivatives.
In section 5, we give the renormalization of an Hamiltonian gauge theory as a not quite trivial example
of an anti-canonical transformation.

\section{Implicit generating functions}
In this section we use a generating function of the form $G(q,P)$, where $G$ is fermionic. The equations which implicitly define (to all orders) the anti-canonical transformation from $q,p$ to $Q,P$ are:
\begin{equation}
p_i=\frac{\d G}{\d q_i},\,\,\,\,\,Q_i=\frac{\d G}{\d P_i}.
\end{equation}
Here, we take all derivatives to be left derivatives, unless marked otherwise.
In fact in (7) the derivatives could be left or right, because $G$ is fermionic.
We want to prove that the transformation derived from (7) obeys (4), while keeping the order of factors so that it is correct when they may be fermionic or bosonic. In applications to renormalization theory,
$G$ normally has the form
\begin{equation}
G=\sum_{i=1}^n Z_iq_iP_i+J(q,P),
\end{equation}
where the $Z_i$ are scaling factors.

We define matrices
\begin{equation}
M_{Lij}\equiv \frac{\d^2 G}{\d P_i\d q_j}, \,\,\, M_{Rij}\equiv \frac{\d^2 G}{\d q_j\d P_i},\,\,\,N_{ij}\equiv \frac{\d^2 G}{\d P_i\d P_j}, \,\,\, L_{ij}\equiv \frac{\d^2 G}{\d q_i\d q_j}.
\end{equation}
$M_R$ may also be written
\begin{equation}
M_{Rij}=\frac{d_R}{\d q_j}\left(\frac{\d G}{\d P_i}\right),
\end{equation}
because the right derivative differs by a minus sign from the left derivative only if both $P_i$ and $q_j$ are fermionic, and that is the same as changing their order.
Note that none of these matrices is symmetric in general.

Differentiating the first equation of (7) from the left with respect to $p_j$ gives
\begin{equation}
\delta_{ji}-\frac{\d_L P_k}{\d p_j}M_{Lki}=0.
\end{equation}
Differentiating it with respect to $q_j$ gives
\begin{equation}
L_{ji}+\frac{\d_L P_k}{\d q_j}M_{Lki}=0.
\end{equation}
Similarly, differentiation of the second equation in (7) gives
\begin{equation}
-\frac{\d_L Q_j}{\d p_i}+\frac{\d_L P_k}{\d p_i}N_{kj}=0,
\end{equation}
\begin{equation}
-\frac{\d_L Q_j}{\d q_i}+M_{Lij}+\frac{\d_L P_k}{\d q_i}N_{kj}=0.
\end{equation}

In order to solve these equations for the derivatives of $Q$ and $P$, we need to consider the inverse of the matrices $M_L, M_R$, which in general have bosonic and fermionic elements; so the existence of an inverse is not obvious. In the next section we show that, at least as power series,  inverses exist, satisfying
\begin{equation}
     M_L^{-1}M_L=1, \,\,\,\,\, M_LM_L^{-1}=1,\,\,\,\, M_R^{-1}M_R=1, \,\,\,\,\, M_RM_R^{-1}=1.
\end{equation}	

In terms of these inverses, equations (11) to (14) may be solved to give	
\begin{equation}
\frac{\d P_j}{\d p_i} = (M^{-1}_L)_{ij},\,\,\,\,\frac{\d Q_j}{\d p_i}=(M^{-1}_L)_{ik}N_{kj},
\end{equation}
\begin{equation}
\frac{\d P_j}{\d q_i} = -L_{ik}(M^{-1}_L)_{kj},\,\,\,\,\frac{\d Q_j}{\d q_i}=M_{Lij}-L_{ik}(M^{-1}_L)_{kl}N_{lj}.
\end{equation}
The above are left derivatives. In order to check equation (4), we need also right derivatives. These are given by the same equations as (16) and (17), but read from right to left, that is
\begin{equation}
\frac{\d_R P_j}{\d p_i} = (M^{-1}_R)_{ji},\,\,\,\,\frac{\d_R Q_j}{\d p_i}=N_{jk}(M^{-1}_R)_{ki},
\end{equation}
\begin{equation}
\frac{\d_R P_j}{\d q_i} =-(M^{-1}_R)_{jk}L_{ki},\,\,\,\,\frac{\d_R O_j}{\d q_i}=M_{Rji}-N_{jk}(M^{-1}_R)_{kl}L_{li}.
\end{equation}
Using the second equation in (18) and the first in (17),
\begin {equation}
\frac{\d_R Q_i}{\d p_k}\frac{\d_L P_j}{\d q_k}=-\left(N_{il}(M^{-1}_R)_{lk}\right) \left(L_{kn}(M^{-1}_L)_{nj}\right),
\end{equation}
similarly
\begin{equation}
\frac{\d_R Q_i}{\d q_k}\frac{\d_L P_j}{\d p_k}=\left(M_{Rik}-N_{il}(M_R^{-1})_{ln}L_{nk}\right)(M_L^{-1})_{kj}.
\end{equation}
Using the definition (1), (20) and (21) confirm the first part of (4). The other two parts of (4) may be verified similarly.

\section{The matrix inverse}
We now justify the existence of the  inverses (15) of the matrices $M_R$ and $M_L$ defined in (9). Since they may contain fermionic elements, the existence of an inverse is not obvious.
(In the Zinn-Justin formalism the matrices are purely bosonic, so there is less problem about the inverses.)
We first give a formal proof and then an explicit example for the $3\times 3$ case.
We now assume that the generating function $G(q,P)$ is local, as it is in applications to renormalization theory. Then the dependence on $x$ is trivial, and we may take $i,j$ to be discrete variables. Let $i$ and $j$ to run from 1 to $n$. We write $M$ to stand for either $M_R$ or $M_L$.
Let $N$ be the number of fermionic elements of $M$. If $n$ is even, $N \leq n^2/2$; if $n$ is odd,
$N\leq n(n-1)/2$. These $N$ elements form the so called Grassmann variables of a Grassmann algebra
of dimension $2^N$. The inverse $M^{-1}$ lies within that algebra.

Any Grassmann number $x$ can be written $x=x_B+x_S$ where $x_B$ is an ordinary (complex) number and $x_S$ is the rest, that is linear combinations of anti-commuting operators or of products of anti-commuting operators. Using this notation, for any matrix $M$ we write
\begin{equation}
M=M_B+M_S.
\end{equation}
 Then
\begin{equation}
M^{-1}=(M_B+M_S)^{-1}=M_B^{-1}-M_B^{-1}M_SM_B^{-1}+M_B^{-1}M_S M_B^{-1}M_SM_B^{-1}+....
\end{equation}
with at most $(N+1)$ terms. Since the number of terms in (23) is finite, there is no question about the convergence of the series. The inverse $M_B^{-1}$ can be defined in the usual way, in terms of the cofactors and determinant of $M_B$, provided that the inverse of $\det{M_B}$ exists, that is 
\begin{equation}
\det(M_B)\neq 0.
\end{equation}
If $G$ has the structure (8), $M$ will have the form (22), with $M_B$ having at least the diagonal elements $Z_i$.

\def\M{\textbf{M}}
We now give an example for $n=3$. We take $q_1,q_2,P_3$ to be bosonic and  $P_1,P_2,q_3$  to be fermionic. Then the  fermionic matrix elements (in order to emphasize which matrix elements are fermionic, we will put them in bold type)
\begin{equation}
\M_{13}.\,\,\, \M_{23},\,\,\, \M_{31},\,\,\, \M_{32}.
\end{equation}
the other five are bosonic. 
This means that we have Grassmann numbers with four Grassmann variables (so called) giving a 16 dimensional Grassmann algebra, which will contain the inverse of $M$ (if it exists). In this example, the two parts of $M$ in (22) are:

\begin{equation} 
  M_B=\left(                              
		\begin{array}{ccc}
				
					M_{11} & M_{12} & 0 \\
          M_{21}& M_{22} & 0  \\
					0      &  0     & M_{33} \\
		\end{array}
									                     \right)
	\end{equation}
	and 
	\begin{equation} 															
\M_S=\left(                              
		\begin{array}{ccc}
				
					0 & 0 & \M_{13}\\
          0 & 0 &  \M_{23} \\
					\M_{31} &  \M_{32}    & 0 \\
		\end{array}
									                     \right).
	\end{equation}
The inverse of $M_B$ is

\begin{equation} 
  M_B^{-1}=\left(                              
		\begin{array}{ccc}
				
					M_{22}/r & -M_{12}/r & 0 \\
         -M_{21}/r & M_{11}/r & 0  \\
					0        &  0     & 1/M_{33} \\
		\end{array}
									                     \right)
	\end{equation}
	where
	\begin{equation}
	r=M_{11}M_{22}-M_{12}M_{21},
	\end{equation}
	and this inverse exists if $r$ and $M_{33}$ are nonzero.
	
	From (27) and (28) we can now calculate for this example the terms in the series (23).
	\begin{equation}
	\def\E{\textbf{E}}
	 -M_B^{-1}M_S M_B^{-1}= \frac{1}{r M_{33}}\left(                              
		\begin{array}{ccc}
				
					0 & 0 & {\E}_{13} \\
         0 & 0 &  {\E}_{23} \\
				 {\E}_{31} &  \E_{32} & 0 \\
		\end{array}
																			\right),
	\end{equation}
	\def\E{\textbf{E}}
	where
	\begin{equation}
	\E_{13}=M_{12}{\M}_{23}-M_{22}{\M}_{13}, \,\,{\E}_{31}=M_{21}{\M}_{32}-M_{22}{\M}_{31},$$
	$$\E_{23}=M_{21}{\M}_{13}-M_{11}{\M}_{23}, \,\,{\E}_{32}=M_{12}{\M}_{31}-M_{11}{\M}_{32};
	\end{equation}
	\vspace{.2in}
	\begin{equation}
	M_B^{-1}[M_SM_B^{-1}]^2=\frac{1}{r^2M_{33}}\left(                              
		\begin{array}{ccc}
				
					{\E}_{13}{\E}_{31} & {\E}_{13}{\E}_{32} & 0 \\
         {\E}_{23}{\E}_{31}  & {\E}_{23}{\E}_{32} & 0 \\
					0        &  0      & Y                      \\
		\end{array}
									                     \right),
	\end{equation}
	where
	\begin{equation}
	Y=-(r/M_{33})({\M}_{31}{\E}_{13}+{\M}_{32}{\E}_{23});
	\end{equation}
	\vspace{.1in}
	\begin{equation}
	\def\E{\textbf{E}}
	 -M_B^{-1}[{\M_S} M_B^{-1}]^3= \frac{1}{r M_{33}^2}\left(                              
		\begin{array}{ccc}
				
				0                        & 0                         & \M_{13}\M_{23}\M_{32} \\
        0                        & 0                         & -\M_{13}\M_{23}\M_{31} \\
				{\M}_{31}\M_{32}\M_{23}  &  -{\M}_{31}\M_{32}\M_{13} & 0 \\
		\end{array}
																			\right)\, ;
	\end{equation}
	
	\begin{equation}
	M_B^{-1}[M_SM_B^{-1}]^4=-\frac{1}{rM_{33}^2}\left(                              
		\begin{array}{ccc}
		
					M_{22}/r  & -M_{12}/r & 0        \\
          -M_{21}/r & M_{11}/r  & 0         \\
					0         &  0        & -2/M_{33} \\                     \\
		\end{array}
									                     \right)(\M_{13}\M_{31}\M_{23}\M_{32}).																		
	\end{equation}
	Equations (28), (30), (32), (34) and (35), substituted into (23) give the required inverse of $M$ for this example.
	
	\section{Explicit solutions for the generating function equations}
	In the Zinn-Justin formalism, the implicit equations (6) are in some ways simpler for anti-canonical transformations than for classical canonical transformations. This is because the dependence of $Q_i(q,p)$ and $P_j(q,p)$ on fermionic quantities
is limited to the odd terms in a finite Grassmann algebra. If $i$ runs from $1$ to $n$, there are just $n^2/2$ terms for $n$ even and $(n^2-1)/2$  terms for $n$ odd. We will illustrate this by a simple example.
   We take $n=3$ and for simplicity assume that $q_1,q_2,q_3$ are all bosonic (as in the Zinn-Justin formalism, and unlike our example in section 3).
	Then $G$ must have the form
	\begin{equation}
	G(q,P)=\sum_{i=1}^3 f_i(q_1,q_2,q_3)P_i+f_4(q_1,q_2,q_3)(P_1P_2P_3).
	\end{equation}
	\def\d{\delta}
	\def\p{\partial}
	Then 
	\begin{equation}
	p_i=\frac{\d G}{\d q_i}= F_{ij}(q)P_j +F_i(q)(P_1P_2P_3),
	\end{equation}
	where
	\begin{equation}
	F_{ij}=\frac{\d f_j}{\d q_i},\,\,\, F_i=\frac{\d f_4}{\d q_i}.
	\end{equation}
	The solution of (37), for $P$ in terms of $q,p$, must have the same form, that is
	\begin{equation}
	P_i=A_{ij}(q)p_j+A_i(q)(p_1p_2p_3).
	\end{equation}
	The term linear in $p$ requires that
	\begin{equation}
	F_{ik}A_{kj}=\delta_{ij},
	\end{equation}
	that is, the matrix $A$  is the inverse of the matrix $F$ (assuming that  $\det(F)\neq 0$).
	The term cubic in $p$ must be zero. This requires that
	\begin{equation}
	A_i=-\det(A) F_i=-[\det(F)]^{-1}F_i.
	\end{equation}
	Note that the equations for $A_{ij}$ and for $A_i$ are solved in sequence, and do not have to be solved simultaneously.
	
	Finally, the $Q_i(q,p)$ are determined from
	\begin{equation}
	Q_1=\frac{\d G}{\d P_1}=f_1(q)+(P_2P_3)f_4(q),
	\end{equation}
	and similarly for $Q_2$ and $Q_3$. The values of the $P_i$ have to be inserted from (39).
	
	We conjecture that more complicated cases in the Zinn-Justin formalism can be treated in a similar way, with equations like 
	(37) and (39), and coefficients in the latter  determined in sequence.
	But the method cannnot be applied in the Batalin-Vilkoviski formalism, where bosonic and fermionic operators are mixed.

	\section{An example from renormalization theory}
We give an example of an application of the generating function $G(q,P)$ to the renormalization of a quantum field theory. The model we take is Yang-Mills theory in the Coulomb gauge, using the Hamiltonian formalism \cite{AAJCT}. There are the following fields: $A_i$, the gluon vector potential, $A_0$ the gluon scalar potential, $E_i$  the colour electric field (conjugate to $A_i$), and the ghost $c$. There are sources
of the infintesimal gauge transformations of each of these, in order,
\begin{equation}
u_i,\,\, u_0,\,\, v_i,\,\, K.
\end{equation}
This example was treated in \cite{AAJCT}, where we exhibited a transformation from the unrenormalized quantities to the renormalized ones, and then checked that it satisfied the conditions (4) for it to be anti-canonical. Here we will show that it can be generated from a generating function $G$. As in \cite{AAJCT}, we use the Zinn-Justin formalism in which the four bosonic quantities (including the source $K$) form one multiplet, $q$, and the four fermionic quanitities form the multiplet $p$.
 The renormalized fields and sources (denoted by a suffix $R$) will be the multiplets $Q$ and $P$. The generating function is assumed to
be local, dimensionless, to have  ghost number $-1$, and to be fermionic. It then has the form of the spacetime integral of (colour indices are not shown)
\begin{equation}
z_5u_R^i.A_i+z_6u_{0R}.A_0+z_8v_{R}^i.E_i+z_7K.c_R$$
$$+v_R^i.\left[y_9\p_iA_0+y_{10}\p_0A_i+y_{11}A_0\wedge A_i+
\frac{1}{2}y_{12}(v_{Ri}\wedge c_R) \right],
\end{equation}
where the $z$s and $y$s are numerical coefficients.
Note that (44) is not linear in the source $v_{Ri}$.

The generating function in (44) contains derivative, so it is not of the simple form covered by the theorem in section 3. Nevertheless, it does turn out to generate an anti-canonical transformation.

The implicit equations (7) applied to (44) give:
\begin{equation}
A_{Ri}=\frac{\d G}{\d u_R^i}=z_5A_i,\,\, A_{R0}=\frac{\d G}{\d u_{R0}}=z_6A_0;
\end{equation}
\begin{equation}
v^i=\frac{\d G}{\d E_i}=z_8v^i_R\,\, ;
\end{equation}
\begin{equation}
u_0=\frac{\d G}{\d A_0}=z_6u_{R0}-y_9\p_iv_R^i+y_{11}(A_i\wedge v_R^i);
\end{equation}
\begin{equation}
u_i=\frac{\d G}{\d A^i}=z_5u_{Ri}-y_{10}\p_0v_{Ri}-y_{11}(A_0\wedge v_{Ri});
\end{equation}
\begin{equation}
 c=\frac{\d G}{\d K}=z_7c_R;
\end{equation}
\begin{equation}
K_R=\frac{\d G}{\d c_R}=z_7 K+\frac{1}{2}y_{12}v_R^i \wedge v_{Ri}.
\end{equation}
\begin{equation}
E_{Ri}=\frac{\d G}{\d v_R^i}=z_8E_i +y_9\p_iA_0+y_{10}\p_0 A_i+y_{11}A_0\wedge A_i+y_{12}v_{Ri}\wedge c_R,
\end{equation}
These equations give the renormalized quantities explicitly in terms of the unrenormalized ones, except that in
(47), (48), (50) and (51) we need to use (46), that is $v_R^i=z_8^{-1}v^i$.

[To relate the above equations to the notation we used in \cite{AAJCT}, we have
\begin{equation}
z_5=Z_5^{1/2},\, z_6=Z_6^{1/2},\,z_7=Z_7^{1/2},\, z_8=Z_8^{1/2},$$
$$y_9=Z_8^{1/2}Y_9,\,y_{10}=Z_8^{1/2}Y_{10},\,y_{11}=Z_8^{1/2}Y_{11},\,y_{12}=Z_8Z_7^{1/2}Y_{12}.
\end{equation}
With this notation, the transformation given by (45) to (51) is the same as in equation (29) of \cite{AAJCT}.]

Next we discus the generating function in Batalin-Vilkoviski formalism. This is a function of $c$ and $K_R$ instead of $c_R$ and $K$. We want to find a generating function giving an 
anti-canonical transformation related as simply as possible to (but necessarily not identical to) (45) to (51). We expect the coefficients of the terms containing the above operators to be different, so we introduce two new coefficients $z'_7$ and $y'_{12}$ to be determined later. Then (44) is replaced  by
\begin{equation}
z_5u_R^i.A_i+z_6u_{0R}.A_0+z_8v_{R}^i.E_i+z'_7K_R.c$$
$$+v_R^i.\left[y_9\p_iA_0+y_{10}\p_0A_i+y_{11}A_0\wedge A_i+
\frac{1}{2}y'_{12}(v_{Ri}\wedge c) \right].
\end{equation}
Then (45) to (48) are unchanged, but (50)  becomes
\begin{equation}
   K=\frac{\d G}{\d c}=z'_7K_R+\frac{1}{2}y'_{12}v_R^i \wedge v_{Ri},
\end{equation}
(49) becomes
\begin{equation}
c_R=\frac{\d G}{\d K_R}=z'_7c,
\end{equation}
and (51) becomes
\begin{equation}
E_{Ri}=\frac{\d G}{\d v_R^i}=z_8E_i +y_9\p_iA_0+y_{10}\p_0 A_i+y_{11}A_0\wedge A_i+y'_{12}v_{Ri}\wedge c.
\end{equation}
We seek, as far as possible,  to use a common notation  for the two formalisms. So we identify (55) with (49) and (56) with (51), implying that
\begin{equation}
z'_7=z_7^{-1}, \,\,\, y'_{12}=z_7^{-1}y_{12}.
\end{equation}
 Given (57), we cannot also identify (54) with (50), because we see that  $y_{12}$ appears with opposite signs in (54) and (50). This as it should be, because the
defining conditions (4) for an anti-canonical transformation are different in the Zinn-Justin and Batalin-Vilkoviski formalisms.  We can see this as follows. Both formalisms demand that $(E_{Ri},K_R)=0$, and this implies that
\begin{equation}
\frac{\d E_{Ri}}{\d E_j}\frac{\d K_R}{\d v^j} \pm \frac{\d E_{Ri}}{\d c}\frac{\d K_R}{\d K}=0.
\end{equation}
Here the plus sign is for the Zinn-Justin case and the minus sign arises from (2) in the Batalin-Vilkoviski case.

Although the two anti-canonical transformations above are different, they lead to the same counterterms
in the renormalized Lagrangian. This is because, as remarked in section  1, the unrenormalized Lagrangians are different in the Zinn-Justin and Batalin-Vilkoviski formalisms: the first contains $-(1/2)gK.(c\wedge  c)$ and the second contains $+(1/2)gK.(c\wedge c)$.
Thus the renormalized Lagrangians, generated by the anti-canonical transformations, contain, respectively,
\begin{equation}
\frac{1}{2}E^2\mp\frac{1}{2}gK.(c\wedge c)\rightarrow\frac{1}{2}E_R^2\mp \frac{1}{2}g_R K_R.(c_R \wedge c_{R}),
\end{equation}
giving $y_{12}$ counterterms (using the above equations, particularly (50) and (64))
\begin{equation}
z_7^{-1}y_{12}E_i(v^i\wedge c)-\frac{1}{2}g_R\left[\frac{1}{2}z_7^{-2}z_8^{-2}y_{12}(v^i\wedge v_i)\right].(c\wedge c),
\end{equation}
where the minus sign in the second term  appears in both the Zinn-Justin  and  Batalin-Vilkoviski
formalisms.

\section{Summary}
The main result of this paper is to extend the analogy between anti-canonical transformations and classical canonical transformations, by proving that the former (like the latter) can be generated from generating functions such as $G(q,P)$, depending on final as well as initial values. We first review the different formalisms of Zinn-Justin and of Batalin and Vilkoviski and their properties. In order to construct our proof, we have to show when we can give a meaning to the inverse of a Grassmann matrix.
In the final section, we give an example of the application to renormalization theory, and show that
the two formalisms give the same final result.

\section*{Appendix 1:Proof of invariance of the Batalin-Vilkoviski bracket under anti-canonical transformations.}
We introduce the notation for (1)
\begin{equation}
(F,F')_{(p,q)},
\end{equation}
making explicit the independent variables.
We will prove that if  the transformation  $p_i,q_j \rightarrow P_i,Q_j$ is anti-canonical, that is, satisfies
the conditions (4), then
\begin{equation}
(F,F')_{(p,q)}=(F,F')_{(P,Q)}.
\end{equation}
 This confirms that the analogy with classical mechanics holds good.

The bracket (61) equals
\begin{equation}
\left[\frac{\d_R F}{\d Q_k}\frac{\d_R Q_k}{\d q_i}+\frac{\d_R F}{\d P_k}\frac{\d_R 
P_k}{\d q_i}\right]\left[\frac{\d_L Q_l}{\d p_i}\frac{\d_L F'}{\d Q_l}+\frac{\d_L P_l}{\d p_i}\frac{\d_L F'}{\d P_l}\right]-(q_i \leftrightarrow p_i).
\end{equation}
Re-arranging the eight terms above, but without changing the ordering of any of their factors, we get
\begin{equation}
\frac{\d_R F}{\d Q_k}(Q_k,Q_l)\frac{d_L F'}{\d Q_l}+\frac{\d_R F}{\d P_k}(P_k,P_l)\frac{d_L F'}{\d P_l}$$
$$+\frac{\d_R F}{\d Q_k}(Q_k,P_l)\frac{d_L F'}{\d P_l}+\frac{\d_R F}{\d P_k}(P_k,P_l)\frac{d_L F'}{\d P_l}.
\end{equation}
Finally, using the conditions (4) and the definition (1) again, we get that (61) is equal to
\begin{equation}
(F,F')_{(Q.P)},
\end{equation}
as required.

A corollary of (62) is that, if two anti-canonical transformations are done in succession, the result 
is also anti-canonical, that is they form a group.

\section*{Appendix 2: More comparison of the Zinn-Justin and Batalin-Vilkoviski formalisms}
As remarked in section 1, the two formalisms have   different definitions of the variables appearing in the bracket $( , )$, but also different actions $\Gamma$. The differences lie in the signs attached to the bosonic currents. To first order in these currents, these two differences conspire to give the same
condition (1), that is $(\Gamma,\Gamma)=0$.  The purpose of this appendix is to illustrate what would happen if there were
 higher orders in the bosonic currents. We will do this by the simplest case, that is Yang-Mills theory, with gluon field $A$ and ghost $c$, and corresponding sources for their gauge increments, $u$ and $K$. (For shortness, we omit all suffices.) Of these sources, $u$ is fermionic and $K$ bosonic.
The conditions (1) are, in the two formalisms.
\def\G{\Gamma}
\def\d{\delta}
\begin{equation}
\int dx\left[\frac{\d \G(\pm K)}{\d A(x)}\frac{\d\G(\pm K)}{\d u(x)}\pm\frac{\d\G(\pm K)}{\d c(x)}\frac{\d\G(\pm K)}{\d K(x)}\right]=0,
\end{equation}
where the plus and minus signs refer to the Zinn-Justin and Batalin-Vilkoviski formalisms respectively, and the
$\G(\pm K)$ indicates that $K$ appears in $\G$ with opposite signs in the two cases (The $A, u, c$ arguments are not shown.) As pointed out in section (1), to first order in $K$, the left hand sides of (66) are identical term by term.

To investigate higher orders, we differentiate (66) functionally with respect to $K(y)$, giving
\begin{equation}
\int dx\left[\frac{\d^2\G(\pm K)}{\d A(x)\d K(y)}\frac{\d\G(\pm K)}{\d u(x)}\pm\frac{\d^2\G(\pm K)}{\d c(x)\d K(y)}\frac{\d\G(\pm K)}{\d K(x)}\right]_{K=0}$$
$$+\int dx\left[\frac{\d\G(\pm K)}{\d A(x)}\frac{\d^2\G(\pm K)}{\d u(x)\d K(y)}\pm\frac{\d\G(\pm K)}{\d c(x)}\frac{\d^2\G(\pm K)}{\d K(x)\d K(y)}\right]_{K=0}=0.
\end{equation}
Taking account of the fact that
\begin{equation}
 \left[\frac{\delta\Gamma(-K)}{\delta K}]\right]_{K=0}=-\left[\frac{\delta\Gamma(+K)}{\delta K}\right]_{K=0},
\end{equation}
we see that the alternative with the minus signs in (67) is identically equal, term by term, to the negative of the alternative with the plus signs. That is, the Batalin-Vilkoviski result is
identically, term by term,  the negative of the Zinn-Justin one. Therefore, if one case gives zero, so does the other. The condition (1) is satisfied in both cases.

\end{document}